\documentstyle[12pt]{article}

\catcode `\@=11
\@addtoreset{equation}{section}

\catcode `\@=12

\newcommand{\be}{\begin{equation}}
\newcommand{\en}{\end{equation}}
\newcommand{\bea}{\begin{eqnarray}}
\newcommand{\ena}{\end{eqnarray}}
\newcommand{\beano}{\begin{eqnarray*}}
\newcommand{\enano}{\end{eqnarray*}}
\newcommand{\bee}{\begin{enumerate}}
\newcommand{\ene}{\end{enumerate}}

\newcommand{\R}{R \!\!\!\! R}
\newcommand{\N}{N \!\!\!\!\! N}
\newcommand{\Z}{Z \!\!\!\!\! Z}

\newcommand{\Lc}{{\cal L}}

\begin{document}

\thispagestyle{empty}
 
\vspace*{1cm}

\begin{center}
{\Large \bf Multi-Resolution Analysis and Fractional Quantum Hall Effect: an Equivalence Result}   \vspace{2cm}\\

{\large F. Bagarello}
\vspace{3mm}\\
  Dipartimento di Matematica ed Applicazioni, 
Fac. Ingegneria, Universit\`a di Palermo, I-90128  Palermo, Italy\\
e-mail: bagarell@unipa.it
\vspace{2mm}\\
\end{center}

\vspace*{2cm}

\begin{abstract}
\noindent 
In this paper we prove that any multi-resolution analysis of $\Lc^2(\R)$  produces, for some values of the filling factor, a single-electron wave function of the lowest Landau level (LLL) which, together with its (magnetic) translated, gives rise to an orthonormal set in the LLL. We also give the inverse construction. Moreover, we extend this procedure to the higher Landau levels and we discuss the analogies and the differences between this procedure and the one previously proposed by J.-P. Antoine and the author.     \end{abstract}

\vspace{2cm}

{\bf PACS Numbers}:  02.30.Nw, 73.43.f

\vfill

\newpage

\section{Introduction}

The role of wavelets in various applications of  mathematics and to some physical problems like signal analysis is now completely established: the existence of a wide literature on this field  is sufficient to give an idea of the amount of people involved in this and related topics. For a clear reading on this subject a standard quotation is \cite{dau}. Reference \cite{wavnew} is an updated book where other interesting aspects of wavelets are discussed. What cannot be found in many textbooks since is still to be understood, is the relevance of wavelets in quantum mechanics: at this moment, in our knowledge, very few are the applications proposed in this field, \cite{antbag,bag1,bag2,antbag2,cho} and \cite{pat} among the others. 

One of the most useful feature of wavelets concerns their  localization properties in both configuration and frequency space. This fact is at the basis of the serie of papers  \cite{antbag,bag1,bag2,antbag2} where different families of orthonormal (o.n.) bases in $\Lc^2(\R)$ are used in the search for the ground state of a two-dimensional elecron gas (2DEG) in an uniform positive background and subjected to an uniform electro-magnetic field. This is the physical system which produces the well known fractional Quantum Hall effect (FQHE). The key fact behind this approach is the existence of an unitary map between $\Lc^2(\R)$ and the lowest Landau level (LLL), that is the subspace of $\Lc^2(\R^2)$ corresponding to the lowest eigenvalue of the free hamiltonian of the 2DEG. This implies that any o.n. basis in $\Lc^2(\R)$ (not necessarely a wavelet one!) produces an o.n. basis in the LLL; for this reason the role of wavelets does not seem so crucial. We will comment again on this approach in section 5.

In this paper we extabilish a deeper connection between wavelets and FQHE. In particular we will show that any multi-resolution analysis (MRA) of $\Lc^2(\R)$ produces {\em automatically} a wave function in $\Lc^2(\R)$ and, as a second step, a wave function in the LLL which turns out to be o.n. to its own (magnetic) translated. This procedure, which works for an even value of the inverse filling factor, is only possible when we start from a MRA, contrarily to what happens in \cite{antbag}, and can also be inverted: to any o.n. basis in the LLL which is generated by a single wave function via the action of magnetic translations  can be associated a MRA.

The paper is organized as follows:

in the next section we quickly review some of the main properties of a MRA and of the kq-representation, \cite{zak}, which turns out to be a technical tool useful to implement the orthonormality condition;

in section 3 we state the problem of orthonormality of the single electron wave functions in connection with the FQHE;

in section 4 we show how, for fillings factors $\nu=\frac{1}{2L}$, $L\in\N$, a MRA produces in a completely natural way a wave function for the 2DEG with the desired orthonormality requirement, and vice-versa;

section 5 is devoted to the comparison between this approach and the one proposed in \cite{antbag}. In particular, the example of the Haar o.n. basis is considered in detail. We also extend our procedure to higher Landau levels;

section 6 contains the conclusions and the plains for the future.

\section{Mathematical tools}

In order to keep the paper self-contained we now quickly review, for reader's convenience, the main properties of the mathematical
tools we will use in the rest of the paper.

\subsection{Multi-Resolution Analysis}

The main result in the theory of MRA is the recipe which allows to construct an orthonormal basis in $\Lc^2(\R)$ starting from a single function $\psi$ and acting on $\psi$ with dilation and translation operators, generating the set:
\be
\{ \psi_{j,k}(x) \equiv 2^{j/2}\psi(2^{j}x-k), j,k \in \Z\}
\label{21}
\en

Such a basis  has the good properties of wavelets,
including space {\em and} frequency localization. This is the key to their
usefulness in many physical and mathematical applications. Let us ow sketch the construction of these o.n. bases of
wavelets. The full story may be found, for instance, in \cite{dau}.

A {\em multi-resolution analysis} of
$L^2(\R)$ is an increasing sequence of closed subspaces 
\be
\ldots \subset V_{-2} \subset V_{-1} \subset V_0 \subset V_1 \subset V_2
  \subset \ldots,
\en
with $ \bigcup_{j \in \Z} V_j$ dense in $L^2(\R)$ and 
$ \bigcap_{j \in \Z} V_j = \{0\}$, and such that 
\begin{itemize}
\item[(1)] 
$f(x) \in  V_j \Leftrightarrow f(2x) \in  V_{j+1}$
\item[(2)] 
There exists a function $\phi \in V_0$, called a  {\em scaling} function, 
such
that  $\{\phi(x-k), k \in {\Z}\}$ is an o.n. basis of $V_0$.
\end{itemize}
Combining (1) and (2), one gets an  o.n. basis of $V_j$, namely 
$ \{\phi_{j,k}(x) \equiv 2^{j/2} \phi(2^jx-k),$ 
$ k \in \Z \}.$ The role of $V_j$  as an approximation space  and in the direct decomposition of $\Lc(\R)$ is discussed in \cite{dau}.

Here we only need to know that the theory asserts the existence of a function $\psi$, called the 
{\em mother} of the wavelets, explicitly computable from $\phi$, such that 
$\{\psi_{j,k}(x) \equiv 2^{j/2} \psi(2^jx-k), j,k \in \Z\}$ 
constitutes an orthonormal basis of $ L^2(\R)$: these are the 
{\em orthonormal wavelets}. 

The construction of  $\psi$ proceeds as follows. First, the inclusion $V_0
\subset V_1$ yields the relation
\be   \label{phi}
\phi(x) = \sqrt{2} \,\sum_{n=-\infty}^\infty \; h_n \phi(2x - n),
\quad h_n = \langle \phi_{1,n} | \phi \rangle.
\en
Taking Fourier transforms, this gives
\be
\widehat\phi(\omega) = m_o(\omega/2) \widehat\phi(\omega/2),   
\label{scal}
\en
where
\be
m_o(\omega) = \frac{1}{\sqrt{2}} \sum_{-\infty}^\infty \; h_n e^{-i n
\omega}
\label{costr1}
\en
is a $ 2\pi$--periodic function. Iterating (\ref{scal}), one gets the
scaling function as the (convergent!) infinite product
\be
\widehat\phi(\omega) = (2\pi)^{-1/2} \prod_{j=1}^\infty \;
m_o(2^{-j}\omega).
\label{costr2}
\en
Then one defines the function $\psi \in W_0 \subset V_1$ by the relation
\be
\widehat\psi(\omega) =  e^{i \omega/2} \; \overline{m_o(\omega/2 + \pi)}
                                          \;  \widehat\phi(\omega/2),
\en
or, equivalently
\be   \label{psi}
\psi(x) = \sqrt{2} \; \sum_{n=-\infty}^\infty \;(-1)^{n-1} h_{-n-1} 
              \phi(2x - n),
\en
and proves that the function $\psi$ indeed generates an o.n. basis with all
the
required properties.

Actually, this procedure does not produce an unique result. Another possibility, which is the one we will use in the example below, gives for the mother wavelet the following expansion:
\be   \label{psi2}
\psi(x) = \sqrt{2} \; \sum_{n=-\infty}^\infty \;(-1)^{n} h_{-n+1} 
              \phi(2x - n),
\en

Various additional conditions may  be imposed on the function $\psi$
(hence on the basis wavelets): arbitrary regularity, several vanishing
moments (in any case, $\psi$ has always mean zero), fast
decrease at infinity, even compact support.   
For instance, $\psi$ has compact support if only finitely many $h_n$ differ
from zero. 

Simple examples of this construction are the Haar basis, which comes  from the scaling function $\phi(x)$  equal to 1 for $0 \leq x < 1$
and 0
otherwise, the  spline functions, \cite{dau}, and so on.

What is more interesting for our purposes is the role of the coefficents $\{h_n\}$ defining the two-scale relation (\ref{phi}). These are complex quantities which, if $\phi(x)$ is normalized, must satisfy the following relation:
\be
\sum_{n\in\Z}|h_n|^2=1.
\label{22}
\en
Furthermore it can be proved using the $2\pi$-periodicity of the function $m_o(\omega)$, together with the ortogonality of the set $\{\phi(x-k)\}$ for $k\in\Z$, that
\be
|m_o(\omega)|^2+|m_o(\omega+\pi)|^2=1 
\label{23}
\en
almost everywhere, \cite{dau}.
This equation can be written in two equivalent forms where the coefficients $h_n$ explicitly appear:
\be
\sum_{n\in\Z}h_n \overline{h_{n+2k}}=\delta_{k,0},\qquad \forall k\in \Z
\label{24}
\en
or
\be
\sum_{n,k\in\Z}h_n \overline{h_{n+2k}}e^{2ik\omega}=1,\qquad \mbox{a.e.}\label{25}
\en
or yet, in a more convenient form,
\be
\frac{1}{2}\sum_{n,l\in\Z}h_n \overline{h_{l}}e^{i(l-n)\omega}(1+(-1)^{l+n})=1,\qquad \mbox{a.e.}\label{25bis}
\en

We end this rapid excursus on MRA with the following remark: the set of coefficients $\{h_n\}$ can be considered as the main ingredient of a MRA since it generates $m_o(\omega)$, $\widehat\phi(\omega)$ and, finally, the mother wavelet $\psi(x)$.

\subsection{kq-representation}

The relevance of kq-representation in many-body physics has been extabilished since its first appearances, \cite{zak}. What was originally a physical tool has became, during the years, also a mathematical interesting object, widely analyzed in the literature, see \cite{dau2,jan} for instance. We give here only few definitions and refere to \cite{zak,jan,zak2} and \cite{bacry} for further reading and for applications.

The genesis of the kq-representation consists in the well known possibility of a simultaneous diagonalization of any two commuting operators. In \cite{zak2} it is shown that the following distributions
\be
\psi_{kq}(x)=\sqrt{\frac{2\pi}{a}} \sum_{n\in\Z}e^{ikna}\delta(x-q-na), \quad\quad k\in[0,a[, \quad q\in[0,\frac{2\pi}{a}[
\label{26}
\en
are (generalized) eigenstates of both $T(a)=e^{ipa}$ and $\tau(\frac{2\pi}{a})=e^{ix2\pi/a}$. Here $a$ is a positive real number which plays the role of a lattice spacing.

How it is discussed in \cite{zak2}, these $\psi_{kq}(x)$ are Bloch-like functions corresponding to infinitely localized Wannier functions. They also satisfy orthogonality and closure properties. This implies that, roughly speaking, they can be used to define a new representation of the wave functions by means of the integral transform $Z: \Lc^2(\R)\rightarrow \Lc^2(\Box)$, where $\Box=[0,a[\times[0,\frac{2\pi}{a}[$, defined as follows:
\be
h(k,q):=(ZH)(k,q):=\int_{\R}d\omega \overline{\psi_{kq}(\omega)}H(\omega),
\label{27}
\en
for all function $H(\omega)\in \Lc^2(\R)$. The result is a function $h(k,q)\in \Lc^2(\Box)$. 

To be more rigorous, $Z$ should be defined first on the functions of ${\cal C}_o^\infty(\R)$ and then extended to $\Lc^2(\R)$ using its continuity, \cite{dau2}. In this way it is possible to give a rigorous meaning to formula (\ref{27}) above.

From now on we will work in the following hypothesis:
\be
a^2=2\pi,
\label{28}
\en
which, also in view of the next section, will correspond to fixing the spacing of the lattice underlying the 2DEG.

Replacing  $\psi_{kq}(x)$ with its explicit expression, formula (\ref{27}) produces
\be
h(k,q)=(ZH)(k,q):=\frac{1}{\sqrt{a}}\sum_{n\in \Z}e^{-ikna}H(q+na),
\label{29}
\en
which can be inverted and gives the $x$-representation $H(\omega)\in \Lc^2(\R)$ of a function $h(k,q)\in \Lc^2(\Box)$ as follows:
\be
H(\omega)=(Z^{-1}h)(\omega)=\int_{\Box}dk \, dq \psi_{kq}(\omega)h(k,q).
\label{210}
\en
Due to  (\ref{26}), this equation gives:
\be
H(x+na)=\frac{1}{\sqrt{a}}\int_0^a dk e^{ikna} h(k,x), \quad \forall x\in [0,a[, \, \forall n\in \Z.
\label{212}
\en

In all the literature concerning kq-representation, the role of the boundary conditions is widely discussed, also in connection with the continuity properties of the functions. For instance, in \cite{boonzak}, a function $h(k,q)\in\Lc^2(\Box)$ is said to be continuous if it is the restriction to the kq-cell of a function continuous in the {\em extended } kq-plane ($k,q\in \R$), and if it satisfies the following boundary conditions
\bea
h(k+a,q)=h(k,q),\nonumber \\
h(k,q+a)=e^{ika}h(k,q),
\label{211}
\ena

which are typical of any function in kq-representation and which will always be assumed here.

\section{Stating the problem}

In this section we will discuss a many-body model of the FQHE looking, in particular,  for the single-electron wave function which generates the ground state of the phisical system in the way described below. This system is simply a two-dimensional electron gas, 2DEG, (that is a gas of electrons constrained in a two-dimensional layer) in a positive uniform background and subjected to an uniform magnetic field along $z$ and an electric field along $y$. 

The hamiltonian of the system can be written
as
\begin{equation}
H^{(N)}=H^{(N)}_0+\lambda(H^{(N)}_c+H^{(N)}_B)\label{31}
\end{equation}
where $H^{(N)}_0$ is the sum of $N$ contributions:
\begin{equation}
H^{(N)}_0=\sum^N_{i=1}H_0(i).\label{32}
\end{equation}
Here $H_0(i)$ describes the minimal coupling of the
electrons with the fields:
\begin{equation}
H_0={1\over 2}\,\left(\underline p+\underline A(r)\right)^2={1\over 2}\,\left(p_x-{y\over 2}\right)^2+{1\over 2}\,\left(p_y+{x\over 2}\right)^2. \label{33}
\end{equation}
Notice that we are adopting here the symmetric gauge $\underline A=\frac{1}{2}(-y,x,0)$ and the same unit as in \cite{bms}.
$H^{(N)}_c$ is the canonical Coulomb interaction between charged
particles:
\begin{equation}
H^{(N)}_c={1\over2}\,\sum^N_{i\not=j}{1 \over|\underline r_i-
\underline r_j|}\label{34}
\end{equation}
and $H^{(N)}_B$ is the interaction of the charges with the
background, whose explicit form can be found in \cite{bms}. 

In the following we will consider, as it is usually done in the literature, $\lambda(H^{(N)}_c+H^{(N)}_B)$ as a perturbation of the free hamiltonian $H^{(N)}_0$, and we will look for eigenstates of $H^{(N)}_0$ in the form of Slater determinants built up single electron wave functions. This approach is known to give good results for low electron (or hole) densities, \cite{bms}. The easiest way to attach this problem consists in
introducing the  new variables
  \be
\label{35}
  P'= p_x-y/2, \hspace{5mm}     Q'= p_y+x/2.
  \en
In terms of $P'$ and $Q'$ the single electron hamiltonian, $H_0$, can be written as 
 \be
\label{36}
  H_{0}=\frac{1}{2}(Q'^2 + P'^2).
  \en
The transformation (\ref{35}) can be seen as a part of a canonical map from $(x,y,p_x,p_y)$ into $(Q,P,Q',P')$ where

   \be
\label{37}
   P= p_y-x/2, \hspace{5mm}    
  Q= p_x+y/2.
   \en
  These operators  satisfy the following commutation relations:
  \be 
\label{38}
 [Q,P] = [Q',P']=i, \quad  [Q,P']=[Q',P]=[Q,Q']=[P,P']=0.
  \en
   It is shown in \cite{daza,mo} that a wave function in the $(x,y)$-space
is
   related to its  $PP'$-expression by the formula
  \be
\label{39}
  \Psi(x,y)=\frac{e^{ixy/2}}{2\pi}\int_{-\infty}^{\infty}\,
  \int_{-\infty}^{\infty}e^{i(xP'+yP+PP')}\Psi(P,P')\,dP dP'.
  \en
  The usefulness of the $PP'$-representation stems from the expression
(\ref{36}) 
  of $H_0$. Indeed, in this representation, the single electron Schr\"{o}dinger equation
admits 
  eigenvectors $\Psi(P,P')$ of $H_0$ of the form $\Psi(P,P')=f(P')h(P)$.
Thus
   the ground state of (\ref{36}) must have the form
 $f_0(P')h(P)$,  where
   \be     \label{310}
  f_0(P')= \pi^{-1/4} e^{-P'^2/2},
  \en
  and the function $h(P)$ is arbitrary, which manifests the degeneracy of
the LLL. With $f_0$ as above formula (\ref{39}) becomes
\be           
 \label{311}
  \psi(x,y) = \frac{e^{ixy/2}}{\sqrt{2}\pi^{3/4}}
\int_{-\infty}^{\infty}\,e^{iyP}e^{-(x+P)^2/2}h(P)\,dP. 
 \en
It is wortwhile to stress that at this stage the Coulomb interaction has not jet been considered (and it will not in this paper!) but the common belief is that the explicit form of $h(P)$ should be fixed by this interaction.

Now the problem arises of how to construct the ground state of the free $N$-electrons hamiltonian $H^{(N)}_0$. We use a suggestion coming from the classical counterpart of this quantum problem. It is very well known that the ground state for  a classical 2DEG is  a (triangular) Wigner crystal: the classical electrons are sharply localized on the sites of a  lattice whose lattice spacing is fixed by the electron density. What we expect, and what was proven in \cite{bms}, is that, at least for certain regions of the filling factor, the quantum ground state should not be very different from this classical picture. Here we only sketch the procedure which is analyzed in more details in \cite{bms,antbag}.

We start introducing the so-called magnetic translation operators
$T(\vec{a_i})$ defined by
   \be
  T(\vec{a_i})\equiv\exp{(i\vec{\Pi}_c\cdot\vec{a_i})}, 
  \;\;i=1,2,
\label{312}
  \en
  where $\vec{\Pi}_c\equiv(Q,P)$ and $\vec{a_i}$ are the lattice
   basis vectors ($\vec{a_1}=a(1,0), \vec{a_2}=\frac{a}{2}(1,\sqrt{3})$ for a triangular lattice).
 
From now on we will work for simplicity in a square lattice with unit cell of area $2\pi$:
\be
\vec{a_1}=a(1,0),\quad \vec{a_2}=a(0,1),\quad\quad a^2=2\pi.
\label{313}
\en
This is not a real limitation and is quite useful to keep the notation simple: moreover, its  generalization to lattices of arbitrary shape is straightforward.

The above {\em rationality} condition on the area has the following useful consequence: 
\be
[T(\vec{a_1}),T(\vec{a_2})]=0.
\label{314}
\en
This is not the only commutativity condition satisfied by the magnetic translations. Due to the commutation relations (\ref{38}), we also find
\be
[T(\vec{a_1}), H_0]=[T(\vec{a_2}),H_0]=0.
\label{314bis}
\en

With the choice (\ref{313}) of the lattice's basis the magnetic translations take a simple form 
\be
T_1:=T(\vec{a_1})=e^{iaQ}, \quad T_2:=T(\vec{a_2})=e^{iaP},
\label{315}
\en
and they act on a generic function $f(x,y)\in \Lc^2(\R^2)$ as follows
\be
f_{m,n}(x,y):=T_1^mT_2^n f(x,y)=(-1)^{mn}e^{i\frac{a}{2}(my-nx)}f(x+ma,y+na).
\label{316}
\en
We see from this formula that, if for instance $f(x,y)$ is localized around the origin, then 
$f_{m,n}(x,y)$ is localized around the lattice site $a(-m,-n)$.

Now we have all the ingredients to construct the ground state of $H^{(N)}_0$ mimiking the classical procedure. We simply start from the single electron ground state of $H_0$ given in (\ref{311}), $\psi(x,y)$. Then we construct a set of copies  $\psi_{m,n}(x,y)$ of $\psi$ as in (\ref{316}), with $m,n\in\Z$. All these functions still belong to the lowest Landau level for any choice of the function $h(P)$ due to (\ref{314bis}). $N$ of these wave functions $\psi_{m,n}(x,y)$ are finally used to construct a Slater determinant for the finite system:
\be
\psi^{(N)}(\underline r_1,\underline r_2,...,\underline r_N)= \!\! \frac{1}{\sqrt{N!}}\left|
\begin{array}{ccccccc}
\psi_{m_1,n_1}(\underline r_1) & \psi_{m_1,n_1}(\underline r_2) & . & . & . & . & \psi_{m_1,n_1}(\underline r_N)  \\ 
\psi_{m_2,n_2}(\underline r_1) & \psi_{m_2,n_2}(\underline r_2) & . & . & . & . & \psi_{m_2,n_2}(\underline r_N)  \\ 
. & . & . & . & . & . & .   \\ 
. & . & . & . & . & . & .   \\
. & . & . & . & . & . & .   \\
\psi_{m_N,n_N}(\underline r_1) & \psi_{m_N,n_N}(\underline r_2) & . & . & . & . & \psi_{m_N,n_N}(\underline r_N)  \\ 
\end{array}
\right|
\label{317}
\en

It is known, \cite{bms}, that in order to get $<\psi^{(N)},\psi^{(N)}>=1$ we need to have \be
<\psi_{m_i,n_i}\psi_{m_j,n_j}>=\delta_{m_i,m_j}\delta_{n_i,n_j}.
\label{317bis}
\en 
In fact, if these translated functions were not o.n., then we would get $\|\psi^{(N)}\|=1+O(N)$, which is obviously divergent for $N$ diverging.
 It is clear, therefore, that if we want to perform easily the thermodynamical limit, ortonormality between differently localized single electron wave functions must be required!

In the rest of this section we will discuss how the requirement (\ref{317bis}) can be handled and, in particular, we will show that the use of kq-representation is quite an useful tool since it produces a very simple constraint. Some of the results we are now going to describe in this section are also due to G. Morchio and F. Strocchi, \cite{bmsprivate}, while the original idea of using kq-representation in connection with an orthonormality constraint is already contained in \cite{bacry} in the proof of completeness of lattice states proposed by the authors.

Let $\psi(x,y)$ be as in (\ref{311}) and 
$\psi_{m,n}(x,y)=T_1^mT_2^n \psi(x,y)=(-1)^{mn}e^{i\frac{a}{2}(my-nx)}\psi(x+ma,y+na).
$
After few computations and using the rationality condition $a^2=2\pi$ we get
\be
\psi_{m,n}(x,y)=\frac{e^{i\frac{xy}{2}+iamy}}{\sqrt{2}\pi^{3/4}}\int_{-\infty}^\infty dP e^{i(y+na)P-(x+ma+P)^2/2}h(P).
\label{318}
\en
We are interested now in finding some conditions on $h(P)$ such that condition (\ref{317bis}), or its equivalent form 
\be
S_{m,n}:=<\psi_{0,0},\psi_{m,n}>=\delta_{m,0}\delta_{n,0},
\label{319}
\en
are satisfied. With the above definitions we find
\be
S_{m,n}=\int_{-\infty}^\infty dp e^{inap}\overline{h(p+ma)}h(p),
\label{320}
\en
which restates the problem of the orthonormality of the wave functions in terms of the PP$'$-representation. In particular we see that, for $m=n=0$, this equation implies that $\psi$ in normalized in $\Lc^2(\R^2)$ if and only if $h(P)$ is normalized in $\Lc^2(\R)$. This reflects the unitarity of the transformation (\ref{39}), which, more in general, implies that any o.n. set in $\Lc^2(\R)$ is mapped in an o.n. set in $\Lc^2(\R^2)$.

In order to use now kq-representation it is convenient to split the integral over $\R$ in an infinite sum of integrals restricted to $[ra,(r+1)a[$, $r\in \Z$, use the kq-representation and then, write everything in terms of a single integral over the unit cell $\Box$. We have, using (\ref{212}) and the well known equality
\be
\sum_{l\in\Z}e^{ixl\frac{2\pi}{c}}=c\sum_{l\in\Z}\delta(x-cl),
\label{321}
\en
$$
S_{m,n}=\sum_{r\in\Z}\int_{ra}^{(r+1)a}dp e^{inap}\overline{h(p+ma)}h(p)=\sum_{r\in\Z} e^{inra^2}\int_0^a dp e^{inap}\overline{h(p+(r+m)a)}h(p+ra)=$$
$$=\sum_{r\in\Z}\frac{1}{a}\int_0^a dq \int_0^a dk \int_0^a dk' e^{ir(k-k')a}e^{inaq-ik'ma}h(k,q)\overline{h(k',q)},
$$
so that
\be
S_{m,n}=\int_\Box dk dq e^{inaq-ikma}|h(k,q)|^2.
\label{322}
\en 
Due to the completeness of the set $\{e^{inaq-ikma}, \,n,m\in\Z\}$ in the unit cell $\Box$, we  conclude that $S_{m,n}=\delta_{m,0}\delta_{n,0}$ if and only if $h(k,q)$ is a phase, so that $|h(k,q)|$ is independent of $k$ and $q$. This result can be considered as a slight generalization of the  procedure discussed in \cite{bacry} to the FQHE for filling factor $\nu=1$. 

It is easy to generalize  this result to a filling $\nu=\frac{1}{2}$. The idea is the following: 

a filling factor $\nu=1$ corresponds to all the sites of our square lattice (of spacing $a=\sqrt{2\pi}$) occupied. A $\nu=\frac{1}{2}$ 2DEG can be seen, on the other way, as if the same lattice was only partially occupied: one lattice site is free and the other is occupied. If we require the orthonormality of the related set of single electron wave functions, it is enough to ask for
$S_{m,2n}=\delta_{m,0}\delta_{n,0}$. This is equivalent also to chosing a different lattice, with an unit cell twice than before and basis vectors $a(1,0)$ and $2a(0,1)$. Of course, we would as well have chosen another lattice with basis vectors  $a(0,1)$ and $2a(1,0)$, or also any other lattice with unit cell of area $4\pi$. We use the first choice just to fix  ideas. Equation (\ref{322}) gives
\be
S_{m,2n}=\int_\Box dk dq e^{i2naq-ikma}|h(k,q)|^2=\delta_{m,0}\delta_{n,0},
\label{323}
\en 
which can be rewritten as
\be
\frac{1}{2}\int_\Box dk dq e^{inaq-ikma}\left(|h(k,\frac{q}{2})|^2+|h(k,\frac{q+a}{2})|^2\right).
\label{324}
\en
This implies, again using the completeness of the functions $e^{inaq-ikma}, \,n,m\in\Z$ in $\Box$, that:
\be
J_2(k,q):=|h(k,\frac{q}{2})|^2+|h(k,\frac{q+a}{2})|^2=\frac{1}{\pi}, \quad \mbox{ almost everywhere for } k,q\in\Box.
\label{325}
\en

The generalization to $\nu=\frac{1}{M}$ is straigtforward: we simply require the orthonormality of the wave functions located at a distance of $M$ sites:  
$$
S_{m,Mn}=\int_\Box dk dq e^{iMnaq-ikma}|h(k,q)|^2=\delta_{m,0}\delta_{n,0}
$$
and, proceding as above, we deduce that $h(k,q)$ must satisfies the equality
\be
J_M(k,q):=|h(k,\frac{q}{M})|^2+|h(k,\frac{q+a}{M})|^2+.....+|h(k,\frac{q+(M-1)a}{M})|^2=\frac{M}{2\pi}, 
\label{326}
\en
almost everywhere for $k,q\in\Box$.

The extension to a filling $\nu=\frac{L}{M}$, with $L$ and $M$ relatively prime, can be performed by imposing that condition $S_{m,n}=\delta_{m,0}\delta_{n,0}$ holds only for those $(m,n)$ corresponding to a square lattice in which only $L$ among $M$ lattice sites are occupied. We will not consider this extention in this paper.

\section{What we get from MRA}

In this section we will describe how two subjects which are so different, at a first sight, as the MRA and the orthonormality condition for a 2DEG discussed previously, are indeed very close.

Let us consider a given MRA of $\Lc^2(\R)$. We have seen in section 2 that to this MRA it is associated a certain set of square-summable complex numbers $\{h_n\}_{n\in\Z}$ satisfying, for instance,  condition (\ref{24}). This set produces a $2\pi$-periodic function $m_o(\omega)$ and, through this, the scaling function $\widehat\phi(\omega)$ and the mother wavelet. 

Now we use the sequence $\{h_n\}_{n\in\Z}$ to define the following function, which strongly reminds $m_o(\omega)$:

\be   \label{41}
  T_2(\omega)=
  \left\{
  \begin{tabular}{ll}
  $\frac{1}{\sqrt{a}}\sum_{l\in\Z}h_le^{-il\omega a},$ &  $\omega\in [0,a[$ \\
  0, & otherwise.
  \end{tabular}
  \right.
\en
It is clear that $T_2(\omega)$ is square integrable and is not periodic. In particular, due to the normalization condition (\ref{22}), we have $\|T_2\|_2^2=\int_{\R}|T_2(\omega)|^2 d\omega=1$. Therefore the kq-transform of this function, $t_2(k,q)=(ZT_2)(k,q)$, is  well defined in $\Lc^2(\Box)$. 

In particular, using (\ref{29}) we find
\be
t_2(k,q)=\frac{1}{\sqrt{a}}\sum_{n\in \Z}e^{-ikna}T_2(q+na).
\label{42}
\en
The boundary conditions (\ref{211}) are obviously satisfied: $t_2(k+a,q)=t_2(k,q)$ and $t_2(k,q+a)=e^{ika}t_2(k,q)$. It is easy to check that $t_2(k,q)$ satisfies also the orthonormality conditions (\ref{325}). In fact, since we are interested to the value of $t_2(k,q)$ only in $\Box$, and since $T_2(\omega)$ is different from zero only for $\omega\in[0,a[$, we conclude that, for $(k,q)\in \Box$,
$$
J_2(k,q)=\frac{1}{a}\left(|T(\frac{q}{2})|^2+|T(\frac{q+a}{2})|^2\right)=\frac{1}{a^2}\sum_{l,s}h_l\overline{h_s}e^{i(s-l)qa/2}(1+(-1)^{l+s})
$$
which is equal to $1/\pi$ a.e. in $k,q\in\Box$, due to (\ref{25bis}). This implies that $t_2(k,q)$ gives rise to a family of functions $\psi_{m,n}(x,y)$ in the LLL mutually orthonormal and corresponding to $\nu=1/2$. We will find the explicit form of these $\psi_{m,n}(x,y)$ in the next section, where we will also compare these results with the ones obtained in \cite{antbag}.

The above procedure can be easily extended to fillings $\nu=\frac{1}{2L}$. The extention to odd denumerator is not so straightforward and will be given elsewhere.

The starting point is again the set $\{h_n\}_{n\in\Z}$, producing a MRA of $\Lc^2(\R)$, satisfying  condition (\ref{24}). Now we define,
\be   \label{43}
  T_{2L}(\omega)=
  \left\{
  \begin{tabular}{ll}
  $\frac{1}{\sqrt{a}}\sum_{l\in\Z}h_le^{-il\omega La},$ &  $\omega\in [0,a[$ \\
  0, & otherwise.
  \end{tabular}
  \right.
\en
Again, this is  a square-integrable functions satisfying $\|T_{2L}\|^2=1$. Defining $t_{2L}(k,q)=(ZT_{2L})(k,q)$ we have, for $k,q\in\Box$, $t_{2L}(k,q) =\frac{1}{\sqrt{a}}T_{2L}(q)=\frac{1}{a}\sum_{l\in\Z}h_le^{-ilq La}$. We also stress that $t_{2L}(k,q)$ satisfies the correct boundary conditions. With these definitions, using the rationality conditions $a^2=2\pi$ and collecting contributions of the form $|t_{2L}(k,\frac{q}{2L})|^2$, $|t_{2L}(k,\frac{q+2a}{2L})|^2$,.... and the 'odd ones', $|t_{2L}(k,\frac{q+a}{2L})|^2$, $|t_{2L}(k,\frac{q+3a}{2L})|^2$,.... we get
\bea
&&\!\!\!\!\!\!\!\!\!\!\!\!J_{2L}(k,q):=|t_{2L}(k,\frac{q}{2L})|^2+|t_{2L}(k,\frac{q+a}{2L})|^2+.....+|t_{2L}(k,\frac{q+(2L-1)a}{2L})|^2=\nonumber\\
&&\!\!\!\!\!\!\!\!\!\!\!\!=L\left( |t_{2L}(k,\frac{q}{2L})|^2+|t_{2L}(k,\frac{q+a}{2L})|^2\right)=\frac{L}{a^2}
\sum_{l,s}h_l\overline{h_s}e^{i(s-l)qa/2}(1+(-1)^{l+s}),
\label{44}
\ena
which is again independent of $k$ and $q$ since it is equal to $L/\pi$ a.e. in $\Box$, due to condition (\ref{25bis}). Finally, equation (\ref{326}) is a consequence of the equality $\nu^{-1}=M=2L$. We conclude that $t_{2L}(k,q)$ produces, in the configuration space, a set of mutually orthonormal wave-functions spanning the LLL for $\nu=\frac{1}{2L}$.

This result, which is in a certain sense rather unexpected because relates two distant fields  as MRA and FQHE, is only  half of the surprise. In fact, in the rest of this section, we will also show that this relation works in the opposite direction. More in details, we will show how to construct, starting from a function $h(k,q)$ which produces an o.n. set of translated functions in the LLL, a set of coefficents $\{h_n\}$ satisfying condition (\ref{25bis}), and, therefore, generating a MRA.

The recipe is rather simple and  requires only few lines: let us suppose to have a function $h(k,q)$ belonging to $\Lc^2(\Box)$, satisfying the  boundary conditions $h(k+a,q)=h(k,q)$, $h(k,q+a)=e^{ika}h(k,q)$ and such that 
\be
|h(k,q/2)|^2+|h(k,(q+a)/2)|^2=\frac{1}{\pi}\quad\quad \mbox{ a.e. in } \Box.
\label{45}
\en
This means that in the configuration space the related set $\{\psi_{m,n}(x,y)\}$ is an o.n. set. 
Let us now define 
\be
h_n(k)=\int_0^adq e^{inaq}h(k,q), \quad\quad k\in [0,a[.
\label{46}
\en
Even if $h_n(k)$ is, in general, a function of $k$ it is straightforward to check that if we take $h(k,q)$ coinciding with $t_2(k,q)$ in (\ref{42}), then $h_n(k)=h_n$ for all $n\in \Z$. This means that the dependence on $k$ may disappear in some relevant situation. It is not so surprising, therefore, to check that
$
\sum_{n\in\Z}h_n(k)\overline{h_{n+2l}(k)}
$
does not depend on $k$ for \underline{any} choice of $h(k,q)$, if the equality (\ref{45}) is satisfied. In fact, using equality (\ref{321}) and condition (\ref{45}), we find
\bea
\sum_{n\in\Z}h_n(k)\overline{h_{n+2l}(k)}=a\int_0^adq |h(k,q)|^2e^{-2ilaq}=&&\nonumber\\
=\frac{a}{2}\int_0^adq e^{-ilaq}(|h(k,q/2)|^2+|h(k,(q+a)/2)|^2)&&\!\!\!\!\!=\frac{a}{2\pi}\int_0^adq e^{-ilaq}=\delta_{l,0}.
\label{47}
\ena
This result shows that any o.n. basis in the LLL for a filling factor $\nu=\frac{1}{2}$ produces a MRA of $\Lc^2(\R)$ which, in general, depends on an external parameter $k\in[0,a[$. The extension to a filling $\nu=\frac{1}{2L}$, $L\in \N$, is straightforward.

We conclude this section remarking that, in view of our results, there exists a complete equivalence between MRA and orthonormality of the single electron wave functions in the LLL (for $\nu=\frac{1}{2L}$, $L\in\N$).

\section{Extension to higher Landau levels and further remarks}

In the first part of this section we analyze the relation between the approach we have discussed here with the one originally proposed in \cite{antbag} and further developped in \cite{bag1,bag2}. In those papers we  used wavelet analysis in connection with the FQHE as we have done here. In \cite{bag2}, in particular, we discussed a toy model suggesting the relevance of single electron wave functions arising from wavelet theory in the construction of a Slater-like ground state for a 2DEG. This construction was carried out in details for the FQHE in \cite{antbag,bag1} using the canonical transformation (\ref{311}) and the PP$'$-representation to generate an o.n. basis of functions in the LLL starting from an o.n. set of wavelets in $\Lc^2(\R)$. This procedure is only apparently close to the one proposed in this paper. The first difference is related to the possibility of extending the approach in \cite{antbag} to \underline{any} o.n. basis of $\Lc^2(\R)$, possibility which does not apply here since the procedure proposed in this paper only works for an o.n. basis generated by a MRA. The second difference concerns the nature of the operators acting on the {\em mother} function which generates the o.n. set in the LLL: in \cite{antbag,bag1} these operators are dilation and translation operators. Here, on the other way, we use the magnetic translations defined in (\ref{312}).

Since, however, these two procedures have something in common, we expect that the  resulting wave functions should not be very different. And, in fact, this is the outcome of this section, where we will explore the details of the easiest example: the Haar wavelet. For this choice the set $\{h_n\}_{n\in\Z}$ reduces to $h_0=h_1=\frac{1}{\sqrt{2}}$, and all the other coefficients are zero. We have shown in \cite{antbag} that this choice produces a function in the LLL localized around the origin which looks like
\be
  H_{00}(x,y) = \frac{e^{-ixy/2}e^{-y^2/2}}{2\pi^{1/4}}\,
\{  2\phi(\frac{x-iy+1/2}{\sqrt{2}})
  -\phi(\frac{x-iy}{\sqrt{2}})-\phi(\frac{x-iy+1}{\sqrt{2}})\},
\label{51}
\en
where $\phi(z):=\frac{2}{\sqrt{\pi}}\int_0^ze^{-t^2}dt$ is the error function, \cite{grad}. The whole set $H_{mn}(x,y)$ is discussed in \cite{antbag}, where also its asymptotic behaviour is discussed in connection with the localization of the electrons. Here we only state the result which will be compared with the one resulting by the approach proposed here. We have
\be
  H_{00}(x,y)
\simeq\frac{e^{ixy/2}e^{-x^2/2}}{2\pi^{1/4}}\sqrt{\frac{2}{\pi}}\,
   \left(\frac{1}{x-iy} +\frac{e^{-1/2-x+iy}}{x-iy+1}  
  -2\frac{e^{-1/8-(x-iy)/2}}{x-iy+1/2}\right),
\label{52}
  \en 
  which displays the Gaussian localization of the wave function 
in the variable $x$ and shows the  rather poor localization in $y$. 

\vspace{3mm}

Let us now proceed in a different way. For a filling $\nu=\frac{1}{2}$ and a generic MRA, the function $T_2$ which produces an o.n. set of translates in the LLL is given in (\ref{41}). Using the transformation rule (\ref{311}) we get
$$
T_2(x,y)=\frac{e^{ixy/2}}{\sqrt{2}\pi^{3/4}}\,\int_{-\infty}^\infty e^{iyQ-(x+Q)^2/2}T_2(Q)= \frac{\sqrt{a}e^{ixy/2}}{2\pi^{3/4}}\,\sum_{l\in\Z}h_l\int_0^ae^{iQ(y-la)-(x+Q)^2/2},
$$
which, for the above choice of coefficients corresponding to the Haar wavelet, gives
\be
T_2(x,y)=\frac{\sqrt{a}e^{ixy/2}}{2^{3/2}\pi^{5/4}}\,\int_0^ae^{iQy-(x+Q)^2/2}(1+e^{-iQa})dQ.
\label{53}
\en
$T_2$ can be written in terms of error function $\phi(z)$ as follows:
\bea
T_2(x,y)=&&\!\!\!\frac{\sqrt{a}e^{-ixy/2-y^2/2}}{4\pi^{3/4}}\,(\phi(\frac{x+a-iy}{\sqrt{2}})
  +\nonumber\\
&&+\phi(\frac{x+a-i(y-a)}{\sqrt{2}})
-\phi(\frac{x-iy}{\sqrt{2}})-\phi(\frac{x-i(y-a)}{\sqrt{2}})),
\label{54}
\ena
whose asymptotic behaviour can be found with the help of \cite{grad}:
\bea
T_2(x,y)\simeq&&\!\!\!\frac{\sqrt{a}e^{+ixy/2-x^2/2}}{2^{3/2}\pi^{5/4}}\,(\frac{1}{x-iy}+\nonumber\\
&&+\frac{e^{\pi-ia(x-iy)}}{x-i(y-a)}  
  -\frac{e^{-\pi-a(x-iy)}}{x+a-iy)}-\frac{e^{-a(x-iy)(1+i)}}{x+a-i(y-a)}).
\label{55}
\ena
This formula shows that, even if the two procedures produces different results, the asymptotic behaviours, that is the localization features of the electrons, coincide for $H_{00}$ and $T_2$. This result can be considered as a consequence of the Balian-Low theorem applied to the present situation, see \cite{dau,antbag2}, and of the Battle theorem for our previous proposal, \cite{antbag,battle,antbag2}. Both these theorems give severe constraints on the localization properties of a wave function when orthonormality constraints of different kind are imposed. We refer to \cite{antbag2} for a rather complete review of the localization problem in a generic Landau level.

\vspace{3mm}

In the last part of this section we extend the orthonormality constraint (\ref{319}) to levels higher than the lowest. 

We begin this analysis with a general remark, which already suggests the final result: orthonormality is required on a set of functions obtained by a single wave function via the action of  
the magnetic translations $T_i$. On the other hand, the passage from a Landau level to the other is obtained with the action of the raising and lowering operators $A'^\dagger$ and $A'$ defined by
\be
A'=\frac{Q'+iP'}{\sqrt{2}},
\label{56}
\en
where $Q'$ and $P'$ are given in (\ref{35}). We have already remarked that the translations $T_i$ commute with $Q'$ and $P'$, and with $A'$ and $A'^\dagger$ as a consequence, so that it is reasonable to expect that the orthonormality constraint does not change very much moving from the lowest to some higher Landau level. This is exactly what happens, as we will now show explicitly for the first excited level. 

All the wave functions of the first Landau level, ILL, are given by formula (\ref{39}) with $\Psi(P,P')=f_1(P')h(P)$. Here $f_1(P')=\frac{\sqrt{2}}{\pi^{1/4}}P'e^{-P'^2/2}$ is the first excited function of the harmonic oscillator. Performing the integration in $P'$ we get
\be
 \label{57}
  \psi(x,y) = \frac{ie^{-ixy/2}}{\pi^{3/4}}
\int_{-\infty}^{\infty}\,e^{iyP}e^{-P^2/2}P h(P-x)\,dP.
 \en
Acting on $\psi(x,y)$ with $T_i$ as in (\ref{316}) and defining $S_{m,n}$ as in (\ref{319}) we get
\beano
&&\!\!\!\!\!\!\!\!\!S_{m,n}=\frac{1}{\pi^{3/2}}\int d^2\underline r
\int_{-\infty}^\infty dp \int_{-\infty}^\infty dp' e^{-ianx-iyp+i(y+na)p'-(p^2+p'^2)/2}pp'\overline{h(p-x)}h(p'-x-ma)=\\
&&\!\!\!\!\!\!\!\!\!=\frac{2}{\sqrt{\pi}}\int_{-\infty}^\infty dx\int_{-\infty}^\infty dq
e^{inaq}\overline{h(q)}h(q-ma)(q+x)^2e^{-(q+x)^2}=\int_{-\infty}^\infty dp e^{inap}\overline{h(p+ma)}h(p),
\enano
which coincides with the  result obtained for the LLL. This means that, when passing to the kq-representation, the wave function originating the o.n. set in the ILL is exactly the same function originating the o.n. set in the LLL. Nedless to say, this does not imply that in the configuration space the two different o.n. sets coincide, because they are generated by different $\psi(P,P')$, belonging to different Landau levels. 

Even if the above result has been obtained only for the ILL, it gives a strong indication that the orthonormality condition in terms of $h(P)$ takes exactly the same form for all the Landau levels. This also follows from our original remark on the commutativity among $T_i$ and $A'^\dagger$.

\section*{Outcome}

In this paper we have proven  a deep connection between a MRA of $\Lc^2(\R)$ and the FQHE. In particular we have shown how a single electron wave function which, together with its magnetic translates, produces an o.n. set in the LLL can be constructed starting from a MRA. This procedure works for $\nu=\frac{1}{2L}$, $L\in\N$. We have also shown that this procedure can be inverted, so that to any o.n. basis of translated functions of the LLL (corresponding to $\nu=\frac{1}{2L}$) can be associated a MRA of $\Lc^2(\R)$. Moreover, we have compared this approach with a similar one, \cite{antbag}, which is close for the final result but is very different for the philosophy. We have finally extended this procedure to other Landau levels. 

What is still to be done is a computation of the energy of the 2DEG for such a basis, in order to see if this procedure can give some hints about the ground state for the FQHE. We also plain to extend this procedure to filling $\nu$ of the form $\nu=\frac{1}{2L+1}$ and, more generally, $\nu=\frac{L}{L'}$, with $L$ and $L'$  relatively prime natural numbers.

\section*{Acknowledgements}

This work has been financially supported by M.U.R.S.T.

\end{document}